\documentclass[aoas,MSNbibl,nameyear,dvips]{arximspdf}
\usepackage{graphicx}
\doi{10.1214/13-AOAS614A} 
\referstodoi{10.1214/12-AOAS614}
\volume{7}
\issue{4}
\pubyear{2013}
\firstpage{1866}
\lastpage{1870}

\makeatletter
\renewcommand{\citep}[1]{(\citeauthor{#1} \citeyear{#1})}
\makeatother

\begin{document}
\begin{frontmatter}

\title{Discussion of ``Estimating the historical and future
probabilities of large terrorist events'' by Aaron Clauset and Ryan Woodard}
\runtitle{Discussion}

\begin{aug}
\author{\fnms{George} \snm{Mohler}\corref{}\ead[label=e1]{gmohler@scu.edu}}
\runauthor{G. Mohler}
\affiliation{Santa Clara University}
\address{Department of Mathematics\\
\quad and Computer Science\\
Santa Clara University\\
Santa Clara, California 95053\\
USA\\
\printead{e1}}
\pdftitle{Discussion of ``Estimating the historical and future
probabilities of large terrorist events'' by Aaron Clauset and Ryan Woodard}
\end{aug}

\received{\smonth{7} \syear{2013}}



\end{frontmatter}

I congratulate \citet{clauset2013woodard} on a very interesting article.
The authors analyze a global terrorism data set with the aim of
quantifying the probability of historical and future catastrophic
terrorism events. Using power law, stretched exponential and log-normal
tail probability models for the severity of events (\#~deaths), the
authors make a convincing argument that a 9/11-sized event is not an
outlier among the catalog of terrorist events between 1968 and 2007.
This study builds upon earlier work by \citet{clauset2007} that I also
recommend for those interested in the statistical modeling of terrorism.

While there is consensus among the models employed by Clauset and
Woodard that 9/11 is not an outlier ($p>0.01$), the estimates are
accompanied by large confidence intervals on how likely a 9/11-sized
event is. In Table~2, where the authors forecast the probability of a
9/11-sized event in 2012--2021, forecasted probabilities range from 0.04
to 0.94 depending on the model and the frequency of events over the time
window. Here the uncertainty has less to \,do with the model
specification and more to do with uncertainty in the frequency of
events over the next decade. Terrorist events do not follow a
stationary Poisson process and the intensity can fluctuate greatly over
a several year period of time.

The authors remark in their discussion that relaxing the stationarity
assumptions and incorporating spatial and exogenous variables may help
tighten the range of forecasted probabilities. I would add here that
some progress has been made, in particular, on modeling terrorist event
time series as nonstationary point processes [see \citet
{porter2012,lewis2012,zammit2012,mohler2013,raghavan2012}]. Terrorism
event processes are history dependent and intensities exhibit
correlations at timescales of weeks and months due to self-excitation
[see \citet{porter2012,lewis2012}] and exogenous effects [see \citet
{raghavan2012,mohler2013}].

\begin{figure}

\includegraphics{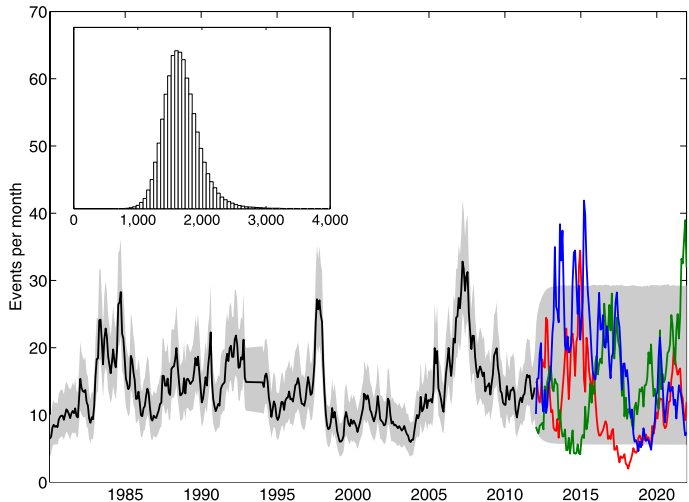}

\caption{Estimated intensity of GTD events ($x_{\mathrm{min}}=10$), posterior
mean (black) and 90\% range (grey), over 1980--2011 and several
forecasted intensities over 2012--2021. Inset figure is the posterior
distribution of the total number of events ($x_{\mathrm{min}}=10$) forecasted
for 2012--2021.}
\label{figure1}
\end{figure}

Log-Gaussian Cox processes (LGCPs) can be used to forecast the
frequency of terrorist events over the next decade, allowing for mean
reversion and some level of smoothness of the intensity. Here we fit
the intensity of a LGCP to the time series of global terrorist attacks
with 10 or more deaths from 1980 to 2011\footnote{Data source: Global
Terrorism Database.} using Langevin Monte Carlo [see \citet{mohler2013}].
The intensity of the process $\lambda_t=e^{x_t}$ is determined by a
Gaussian process $x_t$ satisfying the mean-reverting stochastic
differential equation,
%
%
\begin{equation}
{dx_t}=-\omega(x_t-\mu)\,dt+\alpha \,dB_t.
\end{equation}
We jointly sample the posterior of the intensity and model parameters
and display the posterior mean and 90\% range in Figure~\ref{figure1}.
For each sampled intensity, we use the terminal estimated value at the
end of 2011 and the corresponding posterior parameters to simulate the
intensity forward in time over the range 2012--2021. The inset in Figure~\ref{figure1} displays the posterior distribution of the 10 year
frequency of events over 2012--2021, with most probability mass
contained between 1000 and 3000 events of severity 10 or greater.
This result would indicate that Clauset and Woodard's ``status quo''
forecast is more likely than either the ``optimistic'' or
``pessimistic''
scenarios (it should be pointed out that our forecast of the frequency
of events utilizes GTD data, whereas Clauset and Woodard use RAND-MIPT).

\begin{figure}

\includegraphics{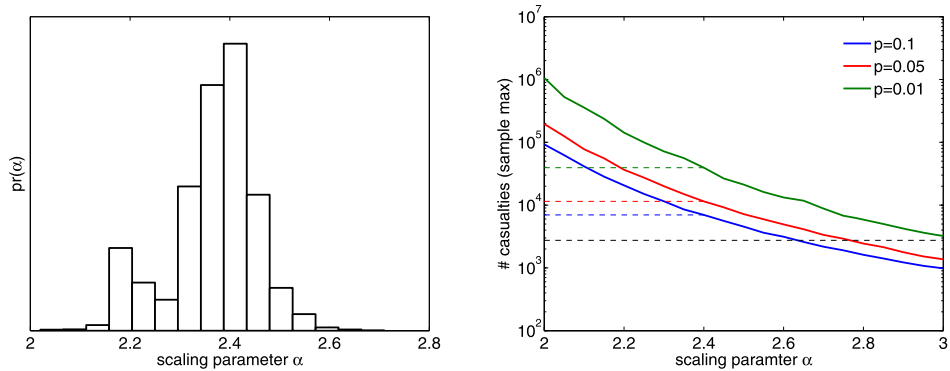}

\caption{Left: Bootstrap probability distribution of estimated scaling
parameter $\alpha$, estimated jointly with $x_{\mathrm{min}}$ from entire
RAND-MIPT data set of 13,274 events. Right: 90th, 95th and 99th
percentiles of the sample max (sample size 994, $x_{\mathrm{min}}\geq10$)
corresponding to varying scaling parameter $\alpha$.}

\label{figure2}
\end{figure}

The probability of catastrophic terrorist events also depends on the
type of weapon used. \citet{clauset2013woodard} find
that the estimated historical probability of a 9/11-sized event is
greatest for explosives ($p=0.37$), fire ($p=0.14$) and firearms
($p=0.12$). Given that 9/11 is categorized as ``other'' and involved a
high degree of planning and coordination, it may not be the case that
all types of terrorism can be modeled alongside ``other'' events as i.i.d.
random variables. In this case it becomes more likely that 9/11 is an
outlier, given that the confidence interval the authors provide for the
historical probability of a catastrophic event of type other is
$[0,0.24]$ and the mean is $p=0.06$.

Ideally, model development should be done in collaboration with domain
experts who can provide insight into whether weapon-dependent severity
probabilities are realistic. Here it may be useful to consider the
distribution of the sample max of competing models to determine
plausibility. In Figure~\ref{figure2} (right) we plot the severity of
the sample max as a function of the scaling parameter for the 90th,
95th and 99th percentiles. For the MLE parameter $\alpha=2.4$, the 95th
percentile of the sample max is approximately 12,000 and the 99th
percentile is approximately 37,000, an order of magnitude larger than
9/11 (dashed line in Figure~\ref{figure2}). The authors note
uncertainty in the estimate for $\alpha$, in particular, 15\% of
bootstrapped estimates cluster around $\alpha=2.2$ (see Figure~\ref{figure2} left).
A decrease in the scaling parameter corresponds to an
increase in the severity threshold separating rare from plausible. The
95th and 99th percentiles for $\alpha=2.2$ correspond to severities of
37,000 and 130,000, respectively. One question that needs to be
addressed is whether events of various types, such as knife or firearm,
can produce an event with $10^4$ or $10^5$ casualties. If not, then the
power law or the i.i.d. assumption may not be appropriate.
\begin{figure}

\includegraphics{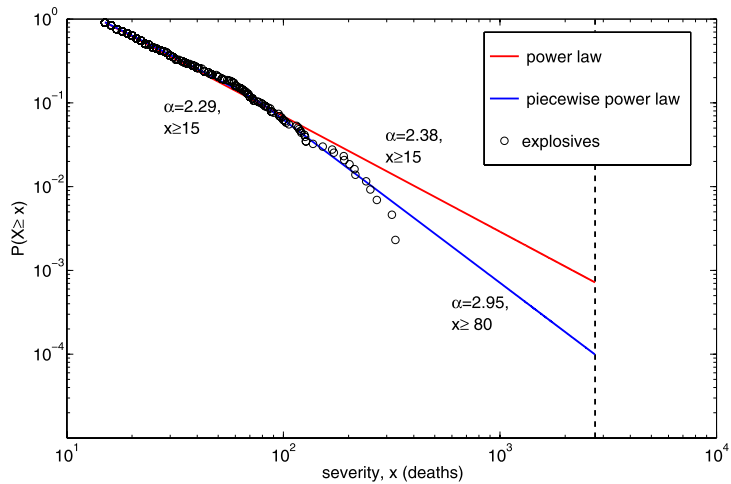}

\caption{Empirical distribution for explosive events along with MLE
power law distribution and piecewise power law distribution.}
\label{figure3}
\end{figure}

We end this discussion by analyzing events of type ``explosive'' that
Clauset and Woodard estimate as having the highest probability
($p=0.37$) among all weapon types of producing a 9/11-sized event. In
Figure~\ref{figure3} we plot the best fit power law model (red) using
the KS estimation procedure Clauset and Woodard outline against the
empirical distribution of events of type explosive. The cutoff point of
the tail $x_{\mathrm{min}}$ is determined by minimizing the $L_{\infty}$
distance between the tail model CDF, $F(x; x_{\mathrm{min}})$, and the empirical
CDF, $G(x;x_{\mathrm{min}})$. The tail model CDF $F$ is estimated for fixed
$x_{\mathrm{min}}$ via MLE and both $F$ and $G$ are normalized to be cumulative
distributions on $[x_{\mathrm{min}},\infty)$.

It should be noted that the power law model exhibits a significant
deviation from the data at the ``extreme'' tail, that is, the $40$--$50$
events with the largest severity. A~weakness of the author's KS
estimation procedure is that, for small $x_{\mathrm{min}}$, error in the extreme
tail is ignored because the KS error satisfies the inequality,
%
%
\begin{equation}
\bigl|F(x;x_{\mathrm{min}})-G(x;x_{\mathrm{min}})\bigr|\leq\max\bigl\{1-F(x;x_{\mathrm{min}}),1-G(x;x_{\mathrm{min}})
\bigr\},\label{KS}
\end{equation}
and the probabilities on the right in (\ref{KS}) are small for large
$x$, small $x_{\mathrm{min}}$ (but may be large for large $x$, large $x_{\mathrm{min}}$).
Thus, the KS procedure may select a value for $x_{\mathrm{min}}$ that fits the
mid-tail over a value that fits the extreme tail. This is not ideal
since the goal here is to estimate the probability of the most extreme
events, not mid-sized events. To give a concrete example, we fit a
piecewise power law to the explosive events and plot the CDF in blue in
Figure~\ref{figure3}. The value $x=80$ is chosen by inspection for the
starting point of the extreme tail, leaving $43$ events to the right. A
likelihood ratio test rejects the power law in favor of the piecewise
power law at the $p=0.07$ level and, furthermore, the piecewise power
law has a lower KS error ($0.027$ compared to $0.033$). The odd property
of the KS estimation procedure is that if the first component of the
piecewise power law is removed, then the model has a worse KS error
than the single power law model even though the extreme tail component
is unchanged. The piecewise power law model corresponds to a historical
9/11 probability of $\approx\!0.04$, an order of magnitude smaller.

We have ignored the role of variance in our discussion of the power law
fit and it is possible that $40$--$50$ events is too small of a sample
size for estimating the tail. A better approach might be to compute the
KS error on the extreme tail only and then choose $x_{\mathrm{min}}$ via
cross-validation. Further research is needed on the selection of
$x_{\mathrm{min}}$, in particular, on how to achieve a good fit in the extreme
tail without overfitting.

%

%



\printaddresses

\end{document}